# Nanocarbons derived from polymers for electrochemical energy conversion and storage – A review

Igor A. Pašti[1]*, Aleksandra Janošević Ležaić[2], Nemanja M. Gavrilov[1], Gordana Ćirić-Marjanović[1], Slavko V. Mentus[1,3]

[1]*University of Belgrade – Faculty of Physical Chemistry, Belgrade, Serbia*

[2]*University of Belgrade – Faculty of Pharmacy, Department of Physical Chemistry and Instrumental Methods, Belgrade, Serbia*

[3]*Serbian Academy of Sciences and Arts, Belgrade, Serbia*

---

* **Corresponding author:** igor@ffh.bg.ac.rs

**Abstract**

Energy demands of modern society require efficient means of energy conversion and storage. Nanocarbons have been identified as versatile materials which combine many desirable properties, allowing them to be used in electrochemical power sources, from electrochemical capacitors to fuel cells. Efficient production of nanocarbons requires innovative and scalable approaches which allow for tuning of their physical and chemical properties. Carbonization of polymeric nanostructures has been demonstrated as a promising approach for production of high-performance nanocarbons with desired morphology and variable surface chemical properties. These materials have been successfully used as active electrode materials in electrochemical capacitors, as electrocatalysts or catalyst supports. Moreover, these materials are often found as parts of composite electrode materials where they play very important role in boosting materials performance. In this contribution we shall review developments in the field of application of polymer-derived nanocarbons for electrochemical energy conversion and storage applications, covering the last decade. Primary focus will be on polyaniline and polypyrrole but carbons derived from other polymers will also be mentioned. We shall emphasize the link between the physical and chemical properties of nanocarbons and their performance in electrochemical power sources with an attempt to derive general guidelines for further development of new materials with improved performances.

**Keywords:** polymer-derived carbons; electrochemical energy conversion; fuel cells; batteries; electrochemical capacitors

# 1. Introduction

Carbon materials have found many applications in electrochemistry. As electrochemical processes take place at the electronic conductor/ionic conductor interface (i.e. electrode) the reasons for such a wide applicability of carbons should be searched in their surface properties. As a rule, carbon materials have highly developed surface with largely developed and complex pore network, possess variety of defects and surface functional groups on the surface and, in addition, are usually highly conductive. The first property is very important because it provides large electrode surface, allowing number of active sites for charge transfer and ion adsorption processes. The presence of defects and surface functional groups is also of great importance as these sites can be active ones for charge transfer, adsorption of reactants and reaction intermediates or anchoring sites for different (nano)particles when it comes to composite formation. Finally, electronic conductivity is crucial for electrochemical applications as in an electrochemical system the charge flow from the outer electronic circuit through the electrode to the electrolyte must not be interrupted. In other words, current collecting properties of carbon electrodes are the reason for their attractiveness in electrochemical applications.

While electrochemistry has many applications in contemporary technologies, by its definition it investigates the conversion of electrical energy into chemical energy and *vice versa*. As such, electrochemistry naturally found a very prominent place in energy conversion applications, which are closely associated with today's energy demands and projected exhaustion of fossil fuels. Modern life cannot be imagined without various types of electrochemical power sources (EPS) which are now used for more than a century. These EPS can be generally classified in three categories: (i) electrochemical capacitors (supercapacitors), (ii) batteries and (iii) fuel cells [1]. The main properties of these EPSs are summarized in Table 1.

**Table 1.** Properties of EPS along with the mechanisms of energy conversion

| EPS | power density / W kg$^{-1}$ | Energy density / Wh kg$^{-1}$ | energy conversion mechanism | process takes place at |
|---|---|---|---|---|
| supercapacitors | $10^3 – 10^7$ | $10^{-2} – 10^{-1}$ | ion adsorption or pseudofaradaic process | electronic conductors/ionic conductor interface |
| batteries | $10^1 –10^3$ | $10^1 – 10^2$ | chemical transformation of electrode material | ideally in entire volume of an electrode material |
| fuel cells | $10^0 – 10^2$ | $10^2 – 10^3$ | (electro)catalytic conversion of reactants | catalytically active sites at the electrode surface |

As can be seen, these three types of EPS cover different energy and power density regimes, which is a consequence of the inherent properties of the processes which underlie the energy conversion mechanism. By considering the mechanisms of energy conversion and the places where the these processes takes place in EPS, one might conclude that the porosity and associated surface are have the most important impact in the case of supercapacitors and fuel cells. In the first case the size of the interface would allow more charge to be stored while in the later case more catalytic sites can be introduced on materials surface. On the other hand, in the case of batteries availability of the surface and possibility to effectively transform the starting material into the product is necessary. Nevertheless, the connections are not so simple, as other physical and chemical properties may also play very important roles in determining an overall performance.

Carbon materials have found application in all three types of EPS. Large specific surface area of carbon materials candidates them for direct application in supercapacitors. However, when it comes to the use of carbon materials in batteries, these are usually parts of composite materials containing various components which undergo chemical transformations during battery operation. While graphite is used in Li-ion batteries due to its intercalation properties [2] this is usually not the case of other types of disordered carbons where proper stacking of graphene layers is lost. Hence, in batteries carbon materials usually play the role of conductive component, or current collector. Concerning fuel cell application, catalytic activity of carbon materials must be considered. In this sense, the most important is the use of carbons as catalysts for oxygen reduction reaction (ORR). Carbon material are, as a rule, active catalysts for ORR in alkaline media [3,4], while there are reports of high catalytic activity in acidic media as well [5]. However, some recent findings do shed a light on these claims as trace level of metals could significantly boost ORR catalytic activity and led to misinterpretation of experimental findings. However, another role of carbons in fuel cell applications is connected with the role of the catalyst support, where finely dispersed catalytically active nanoparticles are dispersed over carbon material, providing uniform distribution, stability and the electrical contact to the outer circuit.

In all the cases carbon properties, their morphology and surface chemistry are extremely important and these can be adjusted by carbon synthesis. While there are numerous strategies to produce carbon materials, carbonization of nanostructured polymers has recently been pointed out as an elegant route to various carbon materials with desired structure and morphology [6-8]. Considering available literature, two polymers are the most common in this type of research: polyaniline (PANI) and polypyrrole (PPy). The conversion of polyaniline and polypyrrole into carbon materials and their application in electrochemistry seems rather natural. First, these two polymeric materials are intensively studied for years and the mechanism of polymerization under various conditions are known, as well as the possibility to tailor the morphology. There are several reviews on the chemistry and application of these materials available [9-12]. However, what we consider to be the most important is the fact that both polyaniline and polypyrrole are conductive polymers which are applied in electrochemistry for quite a while, so, colloquially speaking, there were already at the right place in the right time. Nevertheless, these two polymers are not the only ones used to produce carbons with the application in electrochemical power sources. Our intention here is to present recent advancements in the use of polymers for production of carbons used in electrochemical energy conversion and storage, focusing on the results reported in the last decade. We shall focus on PANI and PPy but other types of polymers used as carbon precursors will also be addressed. Our intension is to provide comprehensive, rather than exhaustive review.

**2. Physical and chemical characterization of carbons – what we know about carbons before putting them in an electrochemical system**

Before proceeding it is important to address the characterization of carbon materials. Taking into the main properties which make these materials attractive for electrochemistry, it is important to probe their specific surface, porosity, morphology, structure (crystal structure and disorder), chemical composition (bulk and surface) and conductivity. The main techniques used for these purposes are summarized in Table 2.

**Table 2.** The main experimental techniques used for characterization of carbon materials

| property | techniques | remarks |
| --- | --- | --- |
| **specific surface and porosity** | BET techniques (gas adsorption measurement) Hg porosimetry | $N_2$ adsorption/desorption $CO_2$ adsorption desorption to probe ultramicropores dry material, measurement starts from vacuum conditions |
| **morphology** | SEM, TEM | analysis under vacuum conditions |
| **crystal structure** | XRD | provides information about bulk properties |
| **structure – presence of defects and functional groups** | Raman spectroscopy | non-destructive, measurements in air |
| | FTIR spectroscopy | measurements in air; typically uses dry materials as water affects spectra |
| **Chemical composition** | elemental analysis | provides information about bulk; determined C, N, O, H and S |
| **Surface chemical composition** | XPS | surface composition; measurement done in vacuum; cannot determine H |
| **Conductivity** | AC conductivity measurements | either pressed in pellet or 4-point probe measurements in thin film configuration |

As can be seen from Table 2, characterization techniques used for determining physical and chemical properties of carbons are *ex situ*. In fact, there is quite a general pattern in papers dealing with carbons electrochemistry. As rule, material is synthesized, the obtained material is characterized using techniques listed in Table 2 (usually not all of them) and then electrochemistry is investigated. Depending on the specific field of application, electrochemical techniques which are used are voltammetry (linear or cyclic, often combined with some hydrodynamic technique like rotating disk electrode, RDE), potentiostatic or galvanostatic measurements and impedance spectroscopy. Finally, electrochemical properties are linked to the results of physical and chemical characterization of materials. As the later is done *ex situ*, the proposed links and justification of electrochemical performance implicitly assumes that the properties of carbon materials are not changed when transferred from, for example, UHV chamber to an electrochemical cell and are potential-independent. Moreover, while in the case of metal electrodes the surface electrochemical processes are well understood and their connection with the electrochemical properties and performance is well established [13,14], this is not the case of carbons. In fact, little is known about possible electrochemical changes of carbon materials under relatively mild conditions typical for operation of different EPS. Hence, a breakthrough in the use of in situ characterization techniques for carbon characterization is needed, and with this in mind we proceed to the overview of the recent results about the use of polymer-derived carbons for in electrochemical power sources.

## 3. Polymer-derived carbons in electrochemical power sources

## 3.1. Electrochemical capacitors

Electrochemical capacitors or supercapacitors can store electrical energy in electrical double layer (EDL) and if this is the only mechanism of charge storage there these are called double layer capacitors. If there are charge transfer processes at the interface these processes can significantly contribute the capacitance of material and in this case the term pseudocapacitor is used [15]. Capacitance of the material is usually reported as gravimetric (specific) capacitance and given in $F\ g^{-1}$, or as areal capacitance and given in $F\ m^{-2}$. In the later case capacitance is evaluated per unit of specific surface area. In the absence of pseudofardaic processes areal capacitance of carbon materials is roughly 10 $\mu F\ cm^{-2}$. Hence, with a carbon having specific surface area of 1000 $m^2\ g^{-1}$, specific capacitance of 100 $F\ g^{-1}$ is easily reached. This is much higher then capacitances of electrostatic capacitors. However, supercapacitors operate at much lower voltage, which is ultimately determined by decomposition voltage of the electrolyte used in the system. With aqueous electrolyte (theoretical decomposition voltage 1.23 V) supercapacitors can operate up to ~1.7 V, but with organic electrolyte or ionic liquid this values is typically between 2.5 and 4 V. This reflects on the energy (W) and power densities (P) of supercapacitors which depend on the capacitance of the material (C), operating voltage (V) and the equivalent series resistance (ESR) as:

$$W = (C \times V^2)/2 \qquad (1)$$

$$P = V^2/(4 \times ESR) \qquad (2)$$

From the equation above it is clear that energy density is limited by materials capacitance and the operating voltage. On the other hand EDL charging and pseudofaradic processes are very fast and the concentrated electrolytes are used so ESR of supercapacitors is very small, typically under 1 . So, operating voltage plays crucial role in determining power densities of supercapacitors.

3.1.1. PANI-based carbons

Direct carbonization of nanostructured PANI results with carbons which are rich in nitrogen, as N is covalently bonded in polymeric precursor. This is very important as the presence of heteroatoms in carbon structure is considered as beneficial for charge storage properties of carbons. During carbonization the morphology of polymeric precursor is typically preserved giving the opportunity to produce carbons with desired morphology. As an example, the work of Yang et al. [16] demonstrated the production of nitrogen-containing carbon nanotubes (N-CNTs) which had open ends and low specific surface. These nanotubes were obtained by carbonization of PANI nanotubes. Authors investigated capacitance of obtained N-CNTs in concentrated KOH solution and found that N-CNTs carbonized at 700 °C exhibit specific capacitance of 163 F g$^{-1}$ at a current load of 0.1 A g$^{-1}$ (Figure 1).

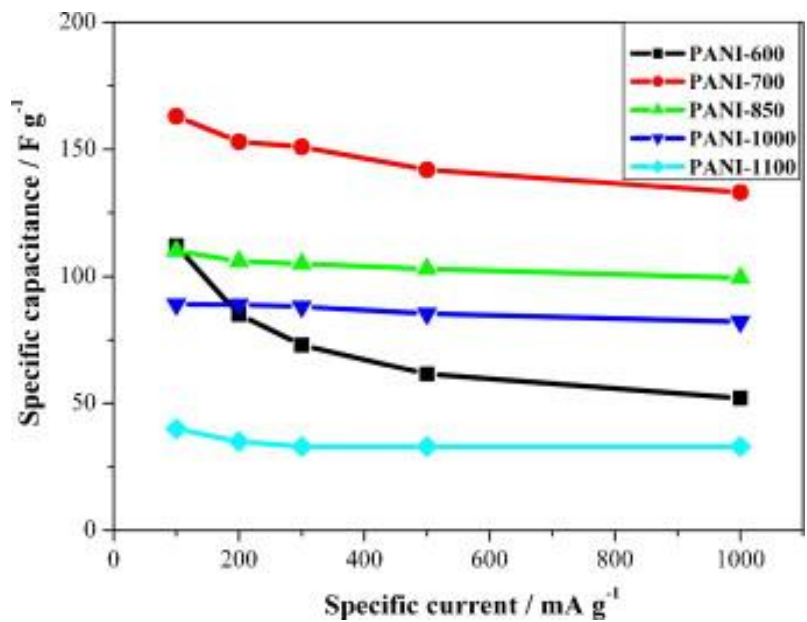

**Figure 1**. The specific capacitances of carbonized PANI nanotubes via different heat treatment temperatures at various specific currents in KOH solution. Reprinted from Ref. [16] with permission. Copyright (2010) Elsevier.

Somewhat higher values of specific capacitance were found by Gavrilov *et al.* [17] who produced a series of nanostructured carbon materials using PANI precursors doped with sulfuric acid, 3,5-dinitrosalicylic acid (DNSA), and 5-sulfosalicylic acid (SSA). Measured capacitances were between 200 and 400 F g−1, estimated from potentiodynamic measurements. The best charge storage properties were ascribed to the balance between highest surface fraction of nitrogen, the highest surface content of pyridinic nitrogen groups, and the highest electrical conductivity, as well as to its well-balanced micro- and mesoporosity and highest content of mesopores, which was found for carbon derived from PANI doped with SSA. Further improvements of capacitive properties of PANI-derived N-containing carbon was demonstrate by the same group [18] using low temperature hydrothermal treatment. Upon the modification of starting C-PANI N/C and O/C ratios were found to increase which led to almost doubled capacitance. Measured capacitances depended on the electrolyte used to assess capacitance and amounted to 363, 220 and 432 F g$^{-1}$, in 6 mol dm$^{-3}$ KOH, 2 mol dm$^{-3}$ KNO3 and 1 mol dm$^{-3}$ H$_2$SO$_4$ (measured potentiodynamically at a scan rate of 5 mV s$^{-1}$). Similar values of specific capacitance of PANI-derived carbons were also found by Wu *et al*. [19] Using 6 mol dm-3 KOH solution for testing, the carbon material produced by carbonization at 600°C displayed specific capacitance of 385 F g$^{-1}$ at the current density of 1 A g$^{-1}$. Xing *et al.* [20] demonstrated the possibility of formation of hollow carbon nanostructures, which is the consequence of the initial morphology of polymeric precursor. Capacitance of 192 F g$^{-1}$ at a current density of 2 A g$^{-1}$ were reported with a rather good cycling stability over 3000 charge/discharge cycles. It is also interesting to mention recent study of Bober *et al*. [21] who produced N,P-containing carbons derived by carbonization of carbonization of conducting polyaniline complex with phosphites. Produced carbons had very low specific surface area (order of 10$^0$ m$^2$ g$^{-1}$) but demonstrated capacitances up to 100 F g$^{-1}$. This led authors to suggest that the surface functionalities formed during the carbonization process have exceptional charge storage properties. It was suggested that further improvements of electrochemical properties can be achieved by enlarging specific surface area with preservation of the type of the surface functional groups. From these reports is it clear that specific surface are does not play the crucial role in the case of capacitive performance of carbon materials. Moreover there are ambiguous claims about the impact of porosity of the capacitive performance. For example, some groups claim that mesoporosity is very important for achieving high capacitance and rate capability of carbons [17], but there are reports which emphasize the impact of microporosity. For example, heteroatom-doped multilocular carbon nanospheres obtained by PANI carbonization at different temperatures, described by Zhou *et al.* [22] are almost completely microporous. These materials display somewhat moderate capacitance (186 F g$^{-1}$) but exceptional rate capability (73.8% capacitance retention at 20 A g$^{-1}$ relative to the capacitance of 0.5 A g$^{-1}$). Some of the highest specific capacitances of PANI derived carbons were demonstrated by Sheng *et al.* [23] who produced high surface area (up to 1645 m$^2$ g$^{-1}$) 3D graphene-sheet like PANI-based carbons. The authors used different transition metals (like Fe, Co, and Ni) which were added during the carbonization process, and catalyzed the graphitization process and served as pore forming agents. The capacitances up to 478 F g$^{-1}$ were measured potentiodynamically (5 mV s$^{-1}$) in KOH solution. All the mentioned studies used aqueous electrolyte so the voltage limitations of such capacitors are rather significant. Using organic electrolyte it is possible to reach higher operating voltages. Sallinas-Torres *et al.* [24] have shown beneficial role of N-functional groups in N-doped activated carbon fibers obtained by carbonization of PANI-modified activated fibers. This specifically relates to enhanced stability of under severe conditions in organic electrolyte, including high charging voltage (3.2 V). The authors ascribed enhanced stability modified carbon fibers to the presence of of aromatic nitrogen group, especially positively charged N-functional groups.

    Besides the use of "pure" PANI-derived carbons, there are examples of composite formation where PANI was converted to carbonaceous form after polymerization in the presence of other components of the composite material. For example, carbonization of PANI polymerized over functionalized multiwalled CNTs (MWNT-COOH) lead to specific capacitances of 149 F g$^{-1}$ [25]. On the other hand,

pyrolysis of PANI nanofibers over carbon cloth [26] led to the specific capacitances of 266 F g$^{-1}$. Al-Enizi *et al.* [27] produced nickel oxide/carbon nanofiber composites by calcinations of Ni(OH)2 deposited over fibers obtained by electrospinning of polyacrylonitrile + PANI + graphene sol-gel mixture. Calcination temperature was found to affect the specific capacitance of final composite material resulting with the maximum capacitance of 738 F g$^{-1}$ for the composite produced at 400 °C.

In addition to the fundamental studies of capacitive properties of different PANI-derived carbons, there are several reports concerning with the construction of realistic electrochemical capacitors. Supercapacitors employing N-doped carbon nanosheets produced by carbonization of PANI nanosheets were demonstrated to have appreciable performance, which depended on the used electrolyte [28]. Using aqueous electrolyte (1 mol dm$^{-3}$ $H_2SO_4$) capacitor delivered 21.6 W h kg$^{-1}$ at a power density of 293.91 W kg$^{-1}$. In solid state (polyvinyl alcohol /$H_2SO_4$) the device delivered 9.8 W h kg$^{-1}$ at 288 W kg$^{-1}$. Symmetric supercapacitor employing carbonized PANI with Fe ions, obtained by polymerization on Ni foam, was demonstrated by Rantho *et al.* [29]. This device used 6 mol dm$^{-3}$ KOH as the electrolyte and operates at 1.65 V, which can be ascribed to poor catalytic activity of C-PANI towards water decomposition. Described supercapacitor delivers somewhat higher energy density of 41.3 W h kg$^{-1}$ with power density of 231.9 W kg$^{-1}$ respectively. The maximum power density was found to be 469.4 kW kg$^{-1}$. Carbonized Fe cations adsorbed on PANI were also used in an asymmetric capacitor as anode [30]. With nickel phosphate/graphene foam composite as the cathode the device operate at voltage up to 1.6 V using 6 mol dm−3 KOH as the electrolyte. The device achieved maximum energy densities of 49 W h kg$^{-1}$ and power density of 499 W kg$^{-1}$, at 0.5 A g$^{-1}$ and had long-term cycling stability. Similar device employing disulfide nanosheets as a cathode was also demonstrated. In KOH electrolyte the device operated up to 1.7 V and exhibited energy and power densities of 27.8 Wh kg$^{-1}$ and 2991.5 W kg$^{-1}$ respectively.

3.1.2. PPy-based carbons

Just like PANI, PPy also contains covalently bonded nitrogen which remains in the carbon structure after carbonization of polymeric precursor. However, there is significantly less reports considering PPy-derived carbons compared to PANI-derived ones.

Carbon nanospheres derived from nanospherical PPy precursor were described Shen *et al*. [31]. Morphology of the precursor was tailored by using 3-chloroperbenzoic acid as oxidant, structure-induced reagent and dopant for polymerization of pyrrole and the morphology of precursor was perfectly preserved after carbonization (Figure 2). Capacitances up to 176 F g$^{-1}$ were measured, which is close to those reported for PANI-based carbons. Similar values of specific capacitances (130 to 210 F g$^{-1}$) were reported for N-doped pororus carbons derived from dopamine-modified PPy [32]. N-containing carbon nanotubes produced by the carbonization of PPy nanotubes (C-PPy-NTs) showed specific capacitance of 220 F g$^{-1}$ at 5 mV s$^{-1}$ and 120 F g$^{-1}$ at 100 mV s$^{-1}$, measured in 6M aqueous potassium hydroxide solution [33]. Gravimetric capacitance was found to decrease in a linear manner with a logarithm of potential sweep rate. Zhou *et al*. [34] reported capacitance values of 272.0 F g$^{-1}$ (5 mV s$^{-1}$) and 105.1 F g$^{-1}$ (100 mV s$^{-1}$) for a similar material, in 1 M sulfuric acid, but it exhibited somewhat higher capacitance fade compared to C-PPy-NTs from ref. [33]. It was shown that upon activation, about 50% increase in capacitance was obtained, with significantly improved capacitance retention, which was attributed to an enlarged surface area and higher O/C and N/C surface ratios [34].

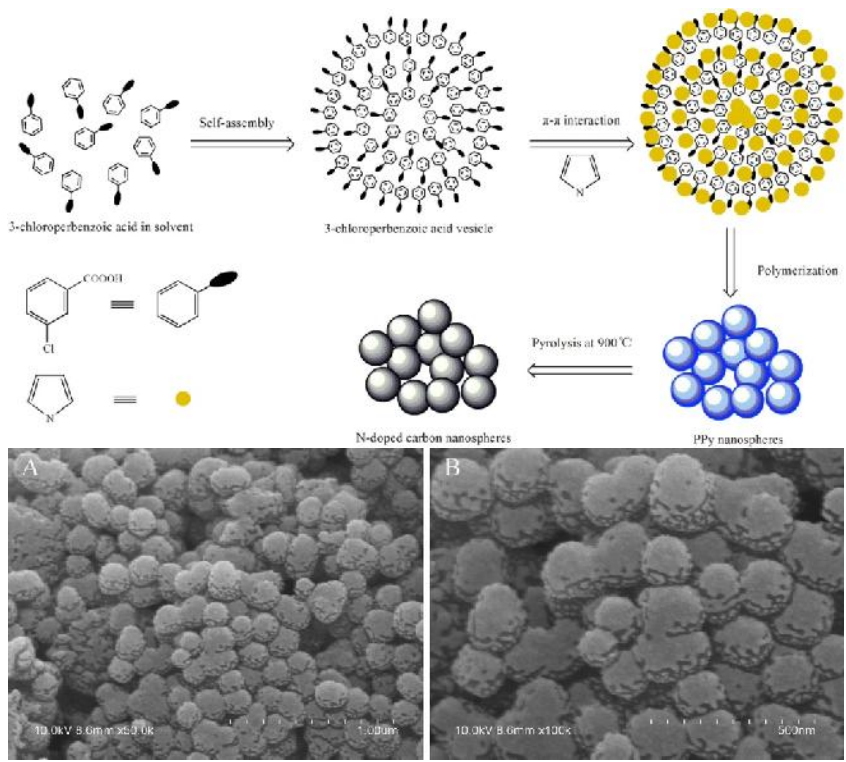

**Figure 2**. Top: Formation mechanism proposed for polypyrrole nanospheres and carbon nanospheres, bottom: SEM images of carbon nanospheres formed from polypyrrole nanospheres. Reprinted from Ref. [31] with permission. Copyright (2014) Elsevier.

There are also several reports concerning composites with carbonized PPy. The work of Chen *et al.* [35] describes production of GO/PPy composite which has very high specific capacitance of 960 F g$^{-1}$. Materials capacitance was found to depend on the PPy content in the composite. Such composite was found to be stable for 300 charge/discharge cycles. Hence, the authors performed thermal treatment of the composite during which GO was reduced and PPy was carbonized. Nanowire morphology of the starting composite material was preserved but the capacitance dropped to 200 F g$^{-1}$. In return, cycling stability and rate capability of the material were significantly improved. PPy derived-carbon composited with metal oxides were also used as materials for supercapacitors like $SnO_2/Co_3O_4$ [36] and $MnO_2$ [37].

3.1.3. Other polymer-derived carbons

There are numerous examples of use of polymeric precursors for production of highly efficient charge storage materials and it is rather difficult to provide complete overview of available results. An immense synthetic potential of various polymeric materials was recently emphasized in the work of Lee *et al.* [38] who produced a number of hypercrosslinked polymers as precursors for heteroatom containing carbons. Besides the fact that some of the produced carbons have exceptionally high capacitances up to 374 F g$^{-1}$, large number of produced materials allowed authors to extract some of very important conclusion regarding the contribution of double layer and pseudocapacitance to the overall capacitive response of heteroatom-containing carbons. It was suggested that pseudocapacitance contribution can be correlated to the surface concentration of heteroatoms (N and S, Figure 3).

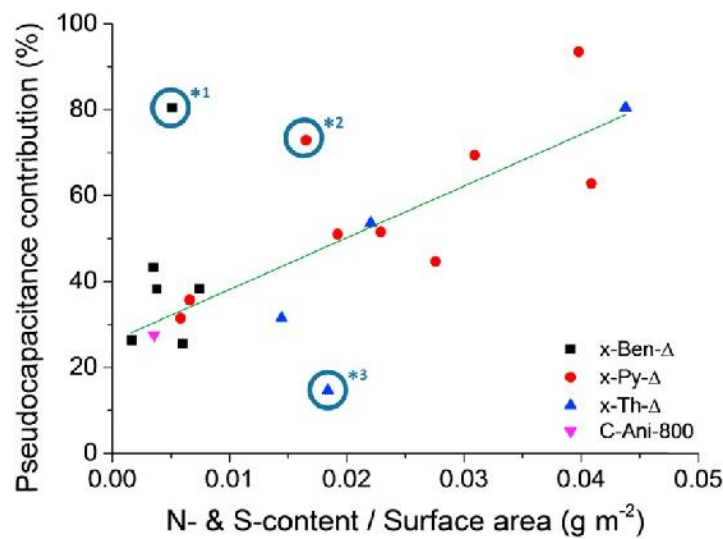

**Figure 3**. Correlation of overall pseudocapacitance contributions against N- & S-content to surface area ratio. *[1] = N-Ben-600. *[2] = N-Py-600. *[3] = N-Th-700. Heteroatom-content is taken from CHNS. Reprinted with permission from Ref. [38].

In Table 3 we summarize some of the results considering polymer-derived carbons for charge storage applications. We have selected materials which displayed high capacitance above 200 F g$^{-1}$.

**Table 3.** Capacitive properties of carbon materials derived from polymers other than PANI and PPy

| Carbon precursor | carbonization temperature / °C | comment | specific surface / m$^2$ g$^{-1}$ | specific capacity / F g$^{-1}$ | reference |
|---|---|---|---|---|---|
| PMMA/PAN core-shell polymer | 800 | KOH activation high oxygen content | 2085 | 314 (@ 0.5 A g$^{-1}$) 237 (@ 20 A g$^{-1}$) | [39] |
| PAN/PMMA | 800 | | 191- 300 | 297.0 | [40] |
| novel polymer obtained from the Schiff base reaction of terephthalaldehyde and m- | 800 – 900 | 850 °C optimal carbonization temperature | 851 (800 °C) 1938 (850 °C) 446 (900 °C) | 397 (@ 0.1 A g$^{-1}$) 145 (@ 100 A g$^{-1}$) | [41] |
| phenylenediamine | | | | | |
| PF resin and PMMA | | average pore size of 4.4 nm | 865 | 252 | [42] |
| P(DVB/VBA) | 800 | KOH activation | 2385 | 319 (@ 0.5 A g$^{-1}$) | [43] |
| PVP | 700 | | 333 | 278.0 | [44] |
| PAN/PMMA | | | 186 | 222.3 | [45] |
| PVDF | 700-1000 | 900 °C optimal carbonization temperature | 1375 | 249 | [46] |

PMMA – polymethylmethacrylate; PAN – polyacrylonitrile; P(DVB/VBA) - phenol-formaldehyde resin (PF) and poly(methyl methacrylate) (PMMA); PF - phenol-formaldehyde; PVP – polyvinylpyrrolidone; PVDF – poly(vinylidene) fluoride

*3.2. Fuel cells and electrocatalysis*

A synonym for fuel cell electrocatalysis is ORR. This reaction attracts attention of electrochemical community for very long time. The overall reactions in acidic and alkaline media are:

$O_2 + 4H^+ + 4e^-$ 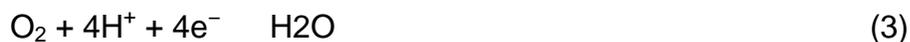 $H_2O$ (3)

and

$O_2 + 2H_2O + 4e^-$ 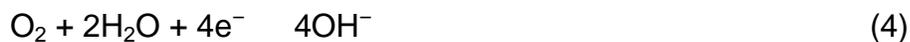 $4OH^-$ (4)

ORR is a common cathode reaction in fuel cell (but oxygen is not the only oxidant used fuel cells, as there are many different types of fuel cell). It is very slow and there is no electrocatalyst on which this reaction is reversible. The best catalysts for ORR are platinum and platinum alloys, but due to its price alternatives are needed. Carbon materials have been demonstrated as possible replacement, at least in alkaline media. In contrast to other ways of carbon synthesis, like catalytic growth over some metal-containing catalyst [8], carbonization of polymeric precursors is an excellent way to avoid any metal contamination of final carbon product, allowing fundamental studies of carbon properties on ORR catalytic performance. However, metals can be intentionally added to polymeric precursor and in such a way introduced in carbon material, which allows boosting catalytic activity. Finally, carbon material can be employed as a catalyst support and due to desirable surface properties can enhance metal anchoring, its stability and dispersion. Carbon decoration with catalytically active nanoparticles also allows the use in many other catalytic reactions for which carbons show no inherent catalytic activity.

3.2.1. PANI-based carbons

Polymerization of aniline is usually performed with oxidants which do not contain any metals hence the obtained polymeric precursor is completely metal-free. If a special care is taken so that there is metal contamination during carbonization process, final carbon materials are completely metal-free. This allows for the fundamental study of catalytic activity of carbon materials and the effects of carbon surface chemistry on catalytic activity. Catalytic activity of different carbonized PANI materials with different morphologies was studied by Ćirić-Marjanović group, showing significant impact of materials morphology, surface functional groups, pororsity and conductivity on ORR activity of C-PANI in alkaline media. Depending on the doping agent (already mentioned sulfuric acid, SSA and DNSA) materials with different morphologies were obtained (nanotubes, nanorods, nanosheets). For example, micro/mesoporous conducting carbonized polyaniline 5-sulfosalicylate nanorods/nanotubes [47] have been identified as very efficient ORR electrocatalysts with high ORR onset potential (above -0.2 V vs. SCE) and high number of electrons consumed per $O_2$ (above 3). In contrast, carbon nanorods obtained by carbonization of self-assembled polyaniline 3,5-dinitrosalicylate nanorods reduced O2 at lower potentials and with lower number of electrons per $O_2$ molecule, suggesting a possible use for hydrogen-peroxide electrosynthesis [48]. A comparison of different C-PANI materials (Figure 4) was provided afterwards [49], suggesting that ORR activity cannot be associated only with the nitrogen content (which was higher than 7 at.%). In fact, it was suggested that the ORR activity is correlated to the concentration of total nitrogen or quaternary nitrogen multiplied by specific surface area. Moreover, correlation between ORR activity and the mesopore surface area was revealed.

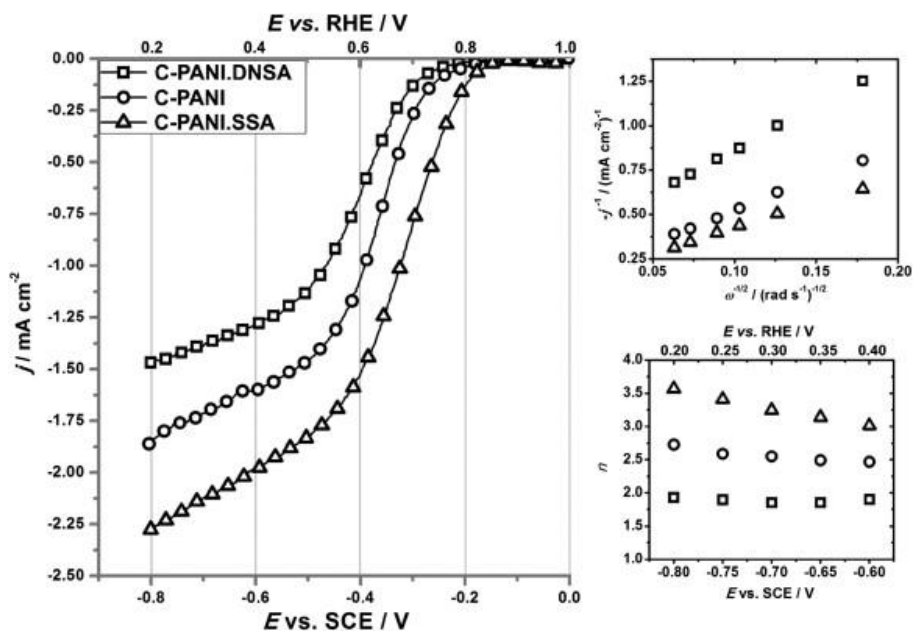

**Figure 4**. Background-corrected RDE voltammograms of oxygen reduction (sweep in anodic direction) on PANI-derived-nanocarbon-modified GC electrodes in $O_2$-saturated 0.1 mol dm$^{-3}$ KOH (loading 250 µg cm$^{-2}$; rotation rate 600 rpm, sweep rate 20 mV s$^{-1}$). Insets are the Koutecky–Levich plots at –0.6 V *vs.* SCE (top right) and the estimated number of electrons (bottom right). Reprinted from Ref. [49] with permission. Copyright (2012) Elsevier.

Further attempts to improve ORR catalytic activity were made by low-temperature hydrothermal treatment of carbonized PANI [50]. When the treatment was performed at 150°C in 1 mol dm$^{-3}$ KOH solution positive shift of onset potential was observed with improved catalyst selectivity at potentials below –0.5 V *vs*. SCE. This effect was ascribed to the alteration of surface functional groups and increase of specific surface. With high catalyst loading ORR onset potential actually matched the one of platinum, while selectivity was also high, suggesting almost complete reduction of $O_2$ to $OH^-$. Moreover, material was found to be stable during extended potentiodynamic cycling in $O_2$-saturated solution and tolerant to ethanol poisoning of the electrolyte. When C-PANI was treated hydrothermally at higher temperature (200°C) [51] further improvements were observed. The effects of the alteration of surface functional groups, which were evidenced by XPS in increased N/C and O/C ratios, was clearly visible in voltammetric tests, unambiguously showing number of surface processes taking place on carbon surface during potential cycling. Both onset potential and the selectivity of C-PANI were improved after hydrothermal treatment. It is important to stress out observed synergistic effect of nitrogen and oxygen surface functional groups observed by Silva *et al*. [52], who particularly emphasized the role of OH surface functional groups. This study was conducted using silica-templated PANI, resulting in mesoporous carbonized product. Described catalysts reached 4e reduction of $O_2$ with activity approaching that of platinum. Another example of ORR catalytic activity improvements upon treatment of PANI-derived carbons can be found in the work of Singh *et al*. [53] who produced N- and S- containing carbonized PANI using carbonization in the presence of iodine.

Interestingly, the authors observed that iodine treatment removes oxygen functional groups and dopant and increases carbon content. Generally, one would expect that such a modification would result in a loss of ORR activity. However, authors found that conductivity is significantly improved, which actually can be correlated to the loss of oxygen and heteroatom functionalities, while specific areas were increased. This increase of the conductivity resulted in increased ORR activity. Hence, obviously, there is no consensus on the impact of different properties of PANI-derived carbons on the ORR activity, even considering the role of heteroatoms, but this seems to be very difficult questions as the results of different groups guide to different conclusions. For the case of monodispersed PANI-based carbons nanoparticles high ORR activity was ascribed to heteroatom dopants and/or intrinsic defect sites and, to some extent, to high porosity and surface area [54]. In terms of the elucidation of the effect of surface functional groups effect on the ORR activity, even more interesting is recent report on P- and N-containing carbons derived from PANI complex with phosphites, This material was mentioned in section 3.1.1. but here we also emphasize very high catalytic activity, considering extremely small specific surface areas [55].

Intentional addition of metals in polymeric precursor is an elegant strategy to develop metal-N co-doped catalysts for ORR. Recently, attention was focused on noble-metal free catalysts with covalently bonded metals to nitrogen incorporated in carbons structure (M-N-C), particularly due to their ORR activity in acidic media. Such catalysts were also prepared using PANI where metal source was added directly to precursor. PANI-derived Fe-N-C catalysts were described by Yan *et al*. [56] and Chen *et al*. [57]. The first group used produced ORR catalysts using PANI/FeCl$_3$ mixtures, while the second one produced catalysts by carbonization of PANI-Prussian blue composite. Very efficient ORR catalysis was observed in both cases in alkaline media (Figure 5). Especially interesting is very high stability of the catalyst which surpasses Pt/C catalyst significantly.

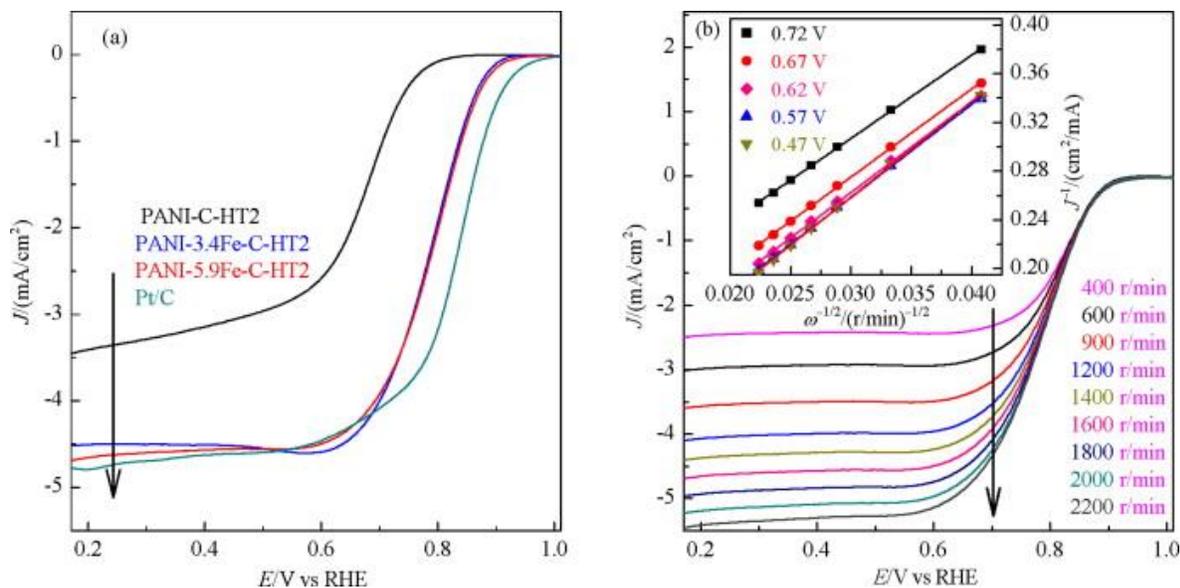

**Figure 5**. (a) Polarization curves of Pt/C (E-TEK, 20%Pt) and PANI-$x$Fe-C-HT2 ($x$ = 0, 4.3, and 5.9 wt%) catalysts in $O_2$-saturated KOH (0.1 mol/L) at a scan rate of 10 mV/s and a rotation rate of 1600 r/min; (b) Polarization curves of the PANI-5.9Fe-C-HT2 catalyst at various rotation rates; the inset shows the corresponding Koutecky-Levich (K-L) plots ($J^{-1}$ vs $\omega^{-1/2}$) at the indicated electrode potentials. Reprinted from Ref. [56] with permission. Copyright (2013) Elsevier.

Considering noble-metal free ORR catalyst metal-oxide-based catalyst attracted large attention. Cheng et al. [58] developed Co-based ORR catalyst which was supported by N-doped PANI-derived carbon hybridized with $TiO_2$. While reaching high ORR activity, demonstrated catalyst also showed improved stability compared to commercial Pt/C and Co/N-doped carbon catalyst. In this case fibrilar PANI precursor was used as precursor for N-doped carbon. As one of the main advantages of this catalyst strong interaction between $TiO_2$ (rutile form) and PANI was outlined. Exceptional ORR activity of mesoporous hybrid shells of carbonized PANI/$Mn_2O_3$ was demonstrated by Cao *et al.* [59]. The results of activity measurement basically matched to that of Pt/C catalyst: very high onset potential of 0.974 V versus Reversible Hydrogen Electrode (RHE), high mass activity and 4-electron reduction of $O_2$. Interestingly, the oxidation state of Mn was the most important in this case, but the effect of C-PANI was also prominent. Pure $Mn_2O_3$ had similar activity as C-PANI, and also C-PANI/$MnO_2$. However, C-PANI/$Mn_2O_3$ catalysts displayed very high activity. Interestingly, $MnO_2$-coated CNTs covered with thin PANI layer and subsequently carbonized also showed ORR activity in acidic media (quite unconventionally measured in HCl solution), but more active catalyst was the one produced at high temperature and with low N content [60]. Considering metal-oxide based catalyst, here we also mention NiO-carbonized PANI composites which are used for methanol oxidation [61,62]. In both reports carbonized PANI was used as a support for NiO and NiO was obtained by calcinations of $Ni(OH)_2$ deposited over pre-carbonized PANI precursor. Strong dependence of specific surface area on the calcinations temperature was reported and resistance to poisoning by reaction intermediates (which is a significant problem in the case of Pt/C catalyst) was very high.

Finally, we address noble metal-based catalysts. In this type of catalysts carbonized PANI is typically used as a support for active noble metal nanoparticles. These nanoparticles are usually deposited on produced carbons and noble metal source is not present in polymeric precursor. We have identified reports dealing with Pt and Pd deposition on carbonized PANI. As carbonized PANI has very rich surface chemistry which can be modified using different chemical agents this is a perfect playground for the investigation of the effects of surface functional groups on metal deposition and, consequently, catalytic activity. Direct deposition of Pt nanoparticles (9 nm in diameter) on carbonized PANI nanotubes/nanosheets was achieved by simple reduction by NaBH$_4$, resulting with a Pt/C-PANI catalysts with high ORR activity in both acidic and alkaline media, and high ethanol oxidation activity in acidic media [63]. Important justification of the of C-PANI as a catalyst support was provided recently by Melke *et al*. [64] who applied a number of different characterization techniques to understand Pt deposition on carbonized PANI, including near-edge X-ray absorption fine structure (NEXAFS). The authors found that fine Pt dispersion over the support with narrow particle distribution can be linked to the presence of pyridinic N-site. In contrast, larger particle sizes and wide particle size distribution is seen for carbons with mainly graphitic and pyrrolic N-groups. It is also important to note that carbonization was found to be beneficial for Pt deposition process and catalytic activity of prepared Pt-based catalysts [65].

Series of Pt/C-PANI materials was developed upon different chemical treatment of C-PANI (NaOH, H$_2$O$_2$ and HNO$_3$) which resulted in modification of surface functional groups of C-PANI and affected Pt nucleation over the support during the reduction with NaBH$_4$ [66]. These catalysts were applied as cathode catalysts in PEM fuel cell and provided improved power density up to 34% compared to the cell employing commercial Pt/C catalysts. It was suggested that surface functional groups introduced by chemical treatment of C-PANI result with reduced particle size, compared to non-treated C-PANI, and more effective ORR catalysis. Wu *et al*. [67] considered that carbonization of PANI at high temperatures, aiming to improve conductivity, increases hydrophobicity of the carbon surface which is not desirable for Pt deposition. The authors grafted sulfuric groups on the surface using sonication in sulfuric acid which significantly reflected of PEM fuel cell performance of Pt catalyst deposited over sulfonated C-PANI. In fact, single cell test confirmed an increase of power and maximum current density with sulfonation level of carbonized PANI. Similar improvements of PEM fuel cell performance were observed upon carboxylation of carbonized PANI nanofibers. Carboxylation was achieved by refluxing carbon in the mixture of sulfuric and nitric acid, and, as a result prepared Pt/C-PANI catalyst displayed better activity than those prepared using Vulcan XC-72 as a catalyst support [68].

Besides Pt, Pd was also deposited on carbonized PANI. While not specifically related to catalytic processes in fuel cells, the work of Koh *et al*. [69] have shown that Pd/C-PANI can efficiently catalyze dehydrogenation of formate and the hydrogenation of carbonate, reversible reactions associated with hydrogen storage and release, and that catalytic activities depend on the concentration of nitrogen in C-PANI. Pd/C-PANI was also used for borohydride oxidation reaction (BOR) in alkaline media as well as for the hydrogen peroxide reduction reaction (HPRR) in acidic media [70].

Finally, there are a few recent examples of the use of PANI-based carbons in bio-catalysis related to the application in bio-fuel cells. Carbonized PANI/graphene oxide hybrid was applied as anode in glucose/O2 fuel cell (Pt/C was used as cathode catalyst) providing a maximum power density of 0.756 mW cm$^{-2}$ at 0.42 V [71]. Yuan *et al*. [72] polymerized aniline over 3D carbon foam and carbonized it to obtain free-standing anode for microbial fuel cell able to deliver up to 1307 mW m$^{-2}$, outperforming commercial carbon felt anodes. Authors identified (i) open porosity, (ii) enlarged electroactive surface, (iii) increased biocompatibility and (iv) decreased electric resistance of the anode scaffold as the reasons for improved performance. A glucose/O$_2$ enzymatic biofuel cell was also constructed using three-dimensional carbonized PANI grown over CNTs [73]. A single cell was able to deliver a maximum power density of 11.2 W m$^{-2}$ at 0.45 V.

3.2.2. PPy-based carbons

Considering PPy-derived carbons used in electrocatalysis, similar pattern like in the case of PANI-based ones can be observed. There are reports investigating catalysis by metal-free carbons, as well as the composites of carbons with metallic and oxide nanoparticles which widen potential applications in the field of electrocatalysis.

Carbonized nanotubular PPy-salt was found to be rather active ORR catalyst in alkaline media [33]. With ORR onset potential close to 0.9 V vs. RHE activity approached that of Pt catalyst but with lower selectivity (around 3 electrons consumed *per* O$_2$ molecule). Similarly to nanostructured PANI-based carbons, high ORR activity was explained by high fraction of mesopores, presence of pyrrolic and pyridynic functional groups and high level of structural disorder, observed by Raman spectroscopy. The conclusion regarding the impact of pyrrolic nitrogen on ORR activity was also derived by Wei *et al*. [74] who produced carbonized polypyrrole-coated graphene aerogel. Nevertheless, the authors also claimed that highly graphitized carbon structures, i.e. highly ordered ones, are responsible for high ORR activity. In the mentioned work, ORR activity was also found to be sensitive on carbonization temperature, with maximum achieved for 600 °C.

Functionalization of PPy-based carbons with metals was achieved either by addition of metal source prior to carbonization or by decoration of carbon materials. High ORR catalytic activity and superioer methanol tolerance of N-doped carbon produced by pyrolysis of $FeCl_3$ and methyl orange (MO) co-doped functional tubular polypyrrole (PPy) was reported by Liu *et al*. [75]. Fe- and Co-doped PPy-based carbons were produced by carbonization of polymeric precursors obtained in the presence of $Fe^{3+}$ and $Co^{2+}$ at 900 °C. Co-doped carbon was more efficient than Fe-doped one, and displayed ORR onset potential in alkaline media similar to that of previously mentioned carbonized PPy-salt nanotubes [76]. However, due to metallic component materials also displayed ORR activity in acidic media, although much lower than Pt/C catalyst. It is interesting that mentioned materials show certain activity in acidic media and both Fe and Co should be easily dissolved in acids, especially under harsh conditions under which ORR takes place. It can be speculated that nitrogen functionalities stabilized metallic component to certain extent and prevented dissolution. A clear role of metal-nitrogen co-doping on ORR activity was recently demonstrated by Sun *et al*. [77] who compared ORR activities of spherical PPy-derived carbons with and without Fe. The catalyst without Fe showed ORR activity only in alkaline media, while Fe-N containing material had excellent ORR activity in both acidic and alkaline media. Moreover, activity in alkaline media matched that of Pt/C catalyst while in acidic media superior poisoning tolerance was revealed (Figure 6). It is a bit surprising that limiting diffusion currents of Pt/C catalyst are lower than that of Fe/N-Cs-900 catalyst (Figure 6) in spite identical selectivity, but this might be due to different catalyst loading and strong bisulfate adsorption on Pt.

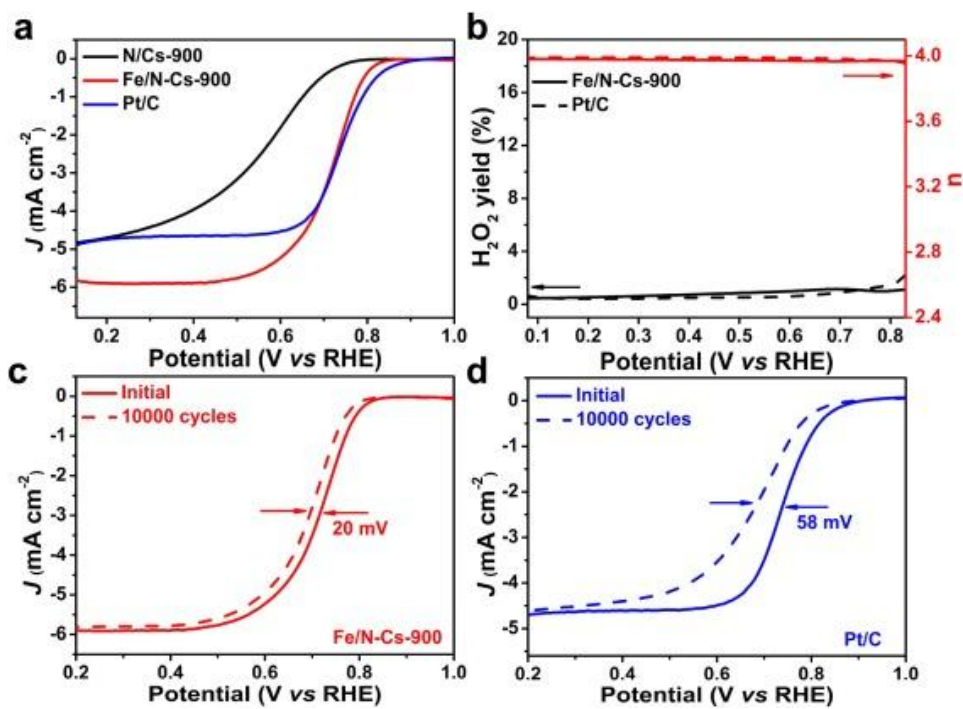

**Figure 6.** (a) ORR polarization plots of N/Cs-900, Fe/N-Cs-900 and Pt/C in $O_2$-saturated 0.5 M $H_2SO_4$. (b) $H_2O_2$ yields and number of electrons consumed *per* $O_2$ molecule (*n*) values of Fe/N-Cs-900 and Pt/C in $O_2$-saturated 0.5 M $H_2SO_4$. (c, d) Endurance tests of Fe/N-Cs-900 (c) and Pt/C (d) for 10 000 cycles in $O_2$-saturated 0.5 M $H_2SO_4$. Reprinted from Ref. [77] with permission. Copyright (2018) Elsevier.

There some more examples of carbonized PPy-based composites with catalytic applications like $Fe_2O_3$/N-doped carbon capsules which were used as support for Pd nanoparticles [78] and carbonized PPy coated $CeNiO_x$ [79]. Although not related to fuel cell electrocatalysis, it is important to mention an interesting approach for decoration of PPy nanotubes with noble metal nanoparticles was demonstrated by Sapurina *et al.* [80]. Authors decorated PPy nanotubes with palladium, platinum, rhodium, or ruthenium nanoparticles, using PPy for reductive deposition of metals. Then the composite was carbonized. While morphology of PPy nanotubes was preserved up to 830 °C, such a high temperature caused metal particles agglomeration. However, at modest temperature (up to 500 °C) PPy was converted to N-doped carbon while metallic nanoparticles were preserved.

3.2.3. Other polymer-derived carbons

In contrast to PANI and PPy, whose inherent properties offer development of versatile carbon materials which can be used as catalyst or a building blocks for the formation of catalytically active composites, other types of polymers are typically used as sacrificial carbon sources for production of composite catalysts. There are many examples of such use of polymeric materials while some of the most recent ones include ORR catalysts derived from ferrocene and melamine/melem [81], Fe-$N_x$-C, B-N and $Fe_3O_4$/$Fe_3C$ containing graphene oxide based catalyst derived from metal-polymer network [82] and Co doped N-containing carbon nanofibers derived from PVP/PAN bicomponent polymers [83]. While the last study falls within the class of M-N containing carbons catalyst, which are known for very high ORR activity, the first two mentioned reports both outline the role of $Fe_3C$ phase for enhanced ORR catalytic activity.

*3.3. Batteries*

Among all the considered ECS, batteries are, at the moment, the most exploited systems for energy conversion and storage. Among them, rechargeable Li-ion batteries cover most of the needs of existing society, including relative fast charging/discharging, high energy density and affordable costs. A breakthrough in Li-ion battery research was discovery of Li intercalation properties of graphite which can be lithiated up to the composition corresponding to $C_6Li$, which is due to layered structure of graphite. Hence carbons which miss graphitic ordering cannot store Li effectively and have low specific capacity. However, Li-intercalate materials which are used Li-ion batteries are usually poor conductors and some conductive component is needed for battery to properly operate. These are typically carbon materials which, due to various morphologies, can offer proper matrix for accommodation of Li-intercalate materials and boost battery performance. Due to limited reserves of Li and increasing demands for energy storage systems there is a tendency for replacement of Li with Na, whose ions are larger and present a challenge for finding an adequate intercalate material. Moreover, there is a constant tendency for development of novel types of batteries. Nowadays, there is a large interest in development of Li-sulfur batteries. These batteries have at least two time higher energy density than Li-ion batteries but the main problem is non conductive nature of sulfur cathode. Namely, during battery discharge series of Li-sulfides is formed and efficient current collector is needed to provide battery operation. An alternative to Li-sulfur batteries are Li-selenium batteries [84], where the main advantages over Li-sulfur batteries are higher electrical conductivity and chemical stability of Li-Se compounds. However, in both cases conductive matrix which will accommodate reactants/product of cathode reaction are needed. Finally, due to exceptional theoretical energy density metal-air, particularly Li-air batteries are actively investigated. This type of batteries has Li-anode and $O_2$-reducing cathode. In order to achieve reversible operation (i.e. to be rechargeable) positive electrode material must possess high ability to reduce $O_2$ to Li-oxygen species ($Li_2O$ or $Li_2O_2$) and also to be able to reduce these species. Hence, in this type of batteries electrocatalytic processes take place on positive electrode, which is in parallel with fuel cells. Hence, while in the case of Li-ion (or in general alkali metal ion batteries) and Li-S and Li-Se batteries carbon serves as conductive component and matrix for accommodation of electrode materials, in the case of $Li-O_2$ batteries the surface of carbon materials incorporated in cathode plays a crucial role through its catalytic action.

3.3.1. PANI-based carbons

There is not so many papers employing carbonized PANI for battery application, but some representative rather recent examples can be found. Xiang *et al.* [85] have used activated carbonized PANI nanotubes as anode material for Li-ion batteries. In principle, material had moderate specific surface are of approx. 620 $m^2\ g^{-1}$ and pore size of 3.10 nm, but delivered exceptionally high first discharge capacity of 1370 mAh $g^{-1}$ and a charge capacity of 907 mAh $g^{-1}$. These capacities were estimated using current load of 100 mA $g^{-1}$. Upon prolonged cycling material reached capacity of 728 mAh $g^{-1}$. Interestingly, XRD analysis have showed diffraction peaks of graphite, but very diffuse, suggesting amorphous material. Authors claimed that such a high capacity of carbonized PANI nanotubes can be ascribed to a high surface area, large porosity, amorphous structure and the high nitrogen- and oxygen-containing functional groups.

Guo *et al.* [86] prepared N-doped carbon composites with $SnO_2$-$Fe_2O_3$ with excellent Li intercalation properties. Li-storage mechanism is based on the reactions of $SnO_2$ and $Fe_2O_3$, but carbon shell formed by PANI carbonization improved electrochemical properties of $SnO_2$-$Fe_2O_3$ significantly (Figure 7). Specific capacities of composite material were found to be between 500 and 1600 mA h kg$^{-1}$, depending on current load. Specific capacitances of $MoS_2$/graphene nanosheets wrapped in carbonized PANI delivered 724 mA h g$^{-1}$ even at a high current density of 1 A g$^{-1}$ [87]. It was suggested that the main role of carbonized PANI is prevention of the pulverization of $MoS_2$ due to the inner-plane volume expansion.

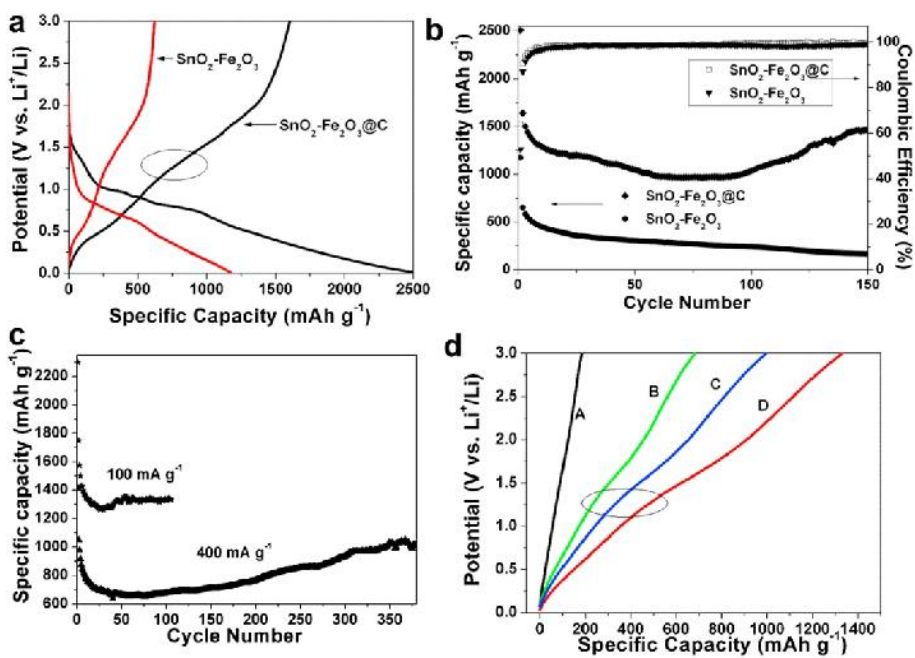

**Figure 7**. (a) The first galvanostatic charge–discharge curve and (b) cyclic performances of $SnO_2$–$Fe_2O_3$ and $SnO_2$–$Fe_2O_3$@C at current density of 200 mA g$^{-1}$. (c) Cyclic performance of $SnO_2$–$Fe_2O_3$@C at current density of 100 mA g$^{-1}$ and 400 mA g$^{-1}$. (d) The 130th galvanostatic charge curve of $SnO_2$–$Fe_2O_3$ at 200 mA g$^{-1}$ (A); the 130th galvanostatic charge curve of $SnO_2$–$Fe_2O_3$@C at 400 mA g$^{-1}$ (B); the 346th galvanostatic charge curve of $SnO_2$–$Fe_2O_3$@C at 400 mA g$^{-1}$ (C); the 130th galvanostatic charge curve of $SnO_2$–$Fe_2O_3$@C at 200 mA g$^{-1}$ (D). Reprinted from Ref. [86] with permission. Copyright (2014) Elsevier.

Caronized PANI/LiFePO$_4$ composite cathode materials were described by Gu *et al.* [88]. Precursor was prepared using *in situ* polymerization of aniline and precipitation of FePO4 in one pot in the presence of hydrogen peroxide as oxidizer. The optimized LiFePO$_4$@C composite contained less than 4 wt.% carbon content, with crystalline pure phase of LiFePO4. A 5 nm thick surface carbon coating layer was formed around LiFePO4 particles. Authors reported capacities between 161.1 and 109.6 mA h g$^{-1}$ for cycling rates between 0.1, and 10 C.

The Na storage capacity of roughly 150 mA h g$^{-1}$ was recently reported for S and N-containing carbon nanotubes produced by carbonization of S-containing PANI nanotubes at high current loads of 5 A g$^{-1}$ [89]. At low current loads capacity of 340 mA h g$^{-1}$ was reported. Authors claimed S incorporation in N-containing carbon nanotubes using ammonium persulfate as both oxidant and sulfur source. It is important to observe much lower sodium storage capacity of nanotubular PANI when compared to Li storage capacity [85]. Liu *et al.* [90] demonstrated the synthesis of amorphous FePO$_4$/carbonized PANI nanorods for sodium storage applications. Material was produced by carbonization of FePO$_4$/PANI precursor and it was outlined that the presence of carbonized PANI enhances electronic transport and improves electrochemical properties. Material was able to deliver Na storage capacity of 140 mA h kg$^{-1}$, which is close to theoretical capacity of NaFePO4 (154.1 mAh g$^{-1}$).

Carbonized PANI based composites have been employed for both Li-S [91] and Li-Se batteries [92]. In both cases electronic conductivity and hollow structure which accommodates polysulfides were outlined as the main factor for excellent performance. It is important to note rather good cycle life of these systems (200 for Li-S and 100 cycles for Li-Se battery).

3.3.2. PPy-based carbons

Pyrolyzed PPy nanofiber webs, prepared at 600°C were employed as anode materials for Li-ion batteries [93]. Depending on the time of pyrolysis (0.5 to 4 h) nitrogen content varied but the highest capacity was observed for material purolysed for 2 hours. The reversible capacity was up to 668 mAh g$^{-1}$ at a current density of 0.1 A g$^{-1}$ and 238 mAh g$^{-1}$ at 5 A g$^{-1}$. Authors claimed that nanofibrous structure and high nitrogen content are responsible for good battery performance of investigated materials. There are also several reports on Li-ion batteries employing carbonized PPy composites with other carbon materials. These include carbon nanofibers derived from polyacrylonitrile and polypyrrole [94], PPy-derived carbon nanospheres grown on graphene [95] and nanotubular carbonized PPy composite with multiwalled CNTs [96]. Many different properties of these materials were outlined as important for effective Li- and Na-storage like high electronic conductivity, small volume changes during intercalation/deintercalation, short ion diffusion paths, easy electrolyte percolation and others. PPy-derived CNTs were used for encapsulation of silicon nanoparticles which are known to suffer from large volume expansion during Li intercalation/deintercalation [97]. This approach led to improved cycling stability of demonstrated electrode materials while specific capacities over 1000 mA h g$^{-1}$ were achieved. Pham-Cong *et al.* [98] described production of SnO2 hollow nanofibers coated with carbonized PPy. N-doped carbon layer was approx. 10 nm thick and significantly contributed Li-storage performance. Namely, after 100 charge/discharge cycles carbon coated $SnO_2$ had capacity of 1648 mA h g$^{-1}$, while bare $SnO_2$ hollow nanofibers deliver capacity of only 386 mA h g$^{-1}$. The authors concluded that improved capacity might be due to the trapping phenomenon of Li$^+$ in N-doped carbon-coated layers and the high efficiency carrier transport effect of the carbon-coated SnO$_2$ hollow nanofibers (Figure 8).

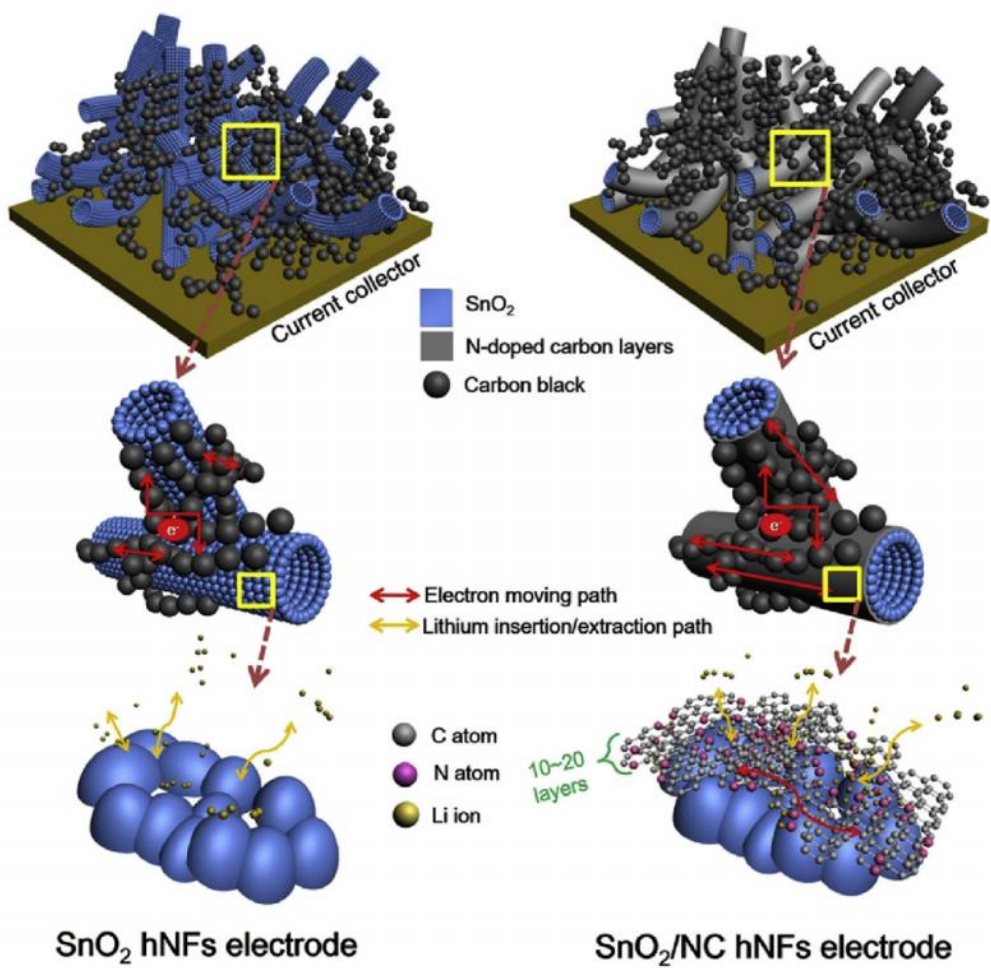

**Figure 8**. Schematic of the lithiation/delithiation processes in both SnO2 hNF and SnO2/NC hNF electrodes. In the case of the SnO2/NC hNFs, several carbon layers containing the Ndoping, defects, and vacancies form on the surface of the SnO2 hNFs. Reprinted from Ref. [98] with permission. Copyright (2017) Elsevier.

The use of different hollow carbonized PPy structures in Li-S batteries led to very high capacities of cathode materials with very good cycling stability. Hollow spherical carbonized PPy/sulfur composite deliver the initial capacity of 1320 mA h g$^{-1}$ and the reversible discharge capacity of 758 mA h g$^{-1}$ after 400 charge/discharge cycles at 0.2 C owing to flexibility of carbonized PPy shells [99]. Quite similar performance was observed for three-dimensional interconnected porous nitrogen-doped graphene/carbonized polypyrrole nanotubes composite [100] and porous N-doped carbon nanotubes derived from tubular PPy [101].

Somewhat lower capacities were observed for KOH activated carbonized PPy nanofibers network which was used in carbon/selenium cathode for Li-Se battery. Initial discharge capacity was 563.9 mAh g$^{-1}$ and the reversible capacity of 414.5 mAh g$^{-1}$ was retained after 100 cycles at 0.2 C charge/discharge rate [102]. High specific surface area and interconnected porous structure were identified as favorable for application of PPy-based carbon as carbon/selenium cathode. Catalytic action of carbonized PPy was employed in Li-O$_2$ batteries with exceptionally high first discharge capacity reaching 8040 mAh g$^{-1}$ at a current density of 0.1 mA cm$^{-2}$ [103]. In the first cycle cell overpotential was only 0.68 V but then increased during the cycling primarily due to the increase of the charge overpotential and in 30$^{th}$ cycle it reached around 1.25 V. upon decoration of carbonized tubular PPy with RuO$_2$ improvements in Li-O$_2$ cell operation, compared to not decorated carbon cathode, were observed [104]. First discharge capacity was as high as 10095 mAh g$^{-1}$ at a current density of 200 mA g$^{-1}$ while good cycle stability was confirmed. Both charge and discharge overpotentials were lower for RuO2-containing cathode than pure carbonized PPy cathode (Figure 9).

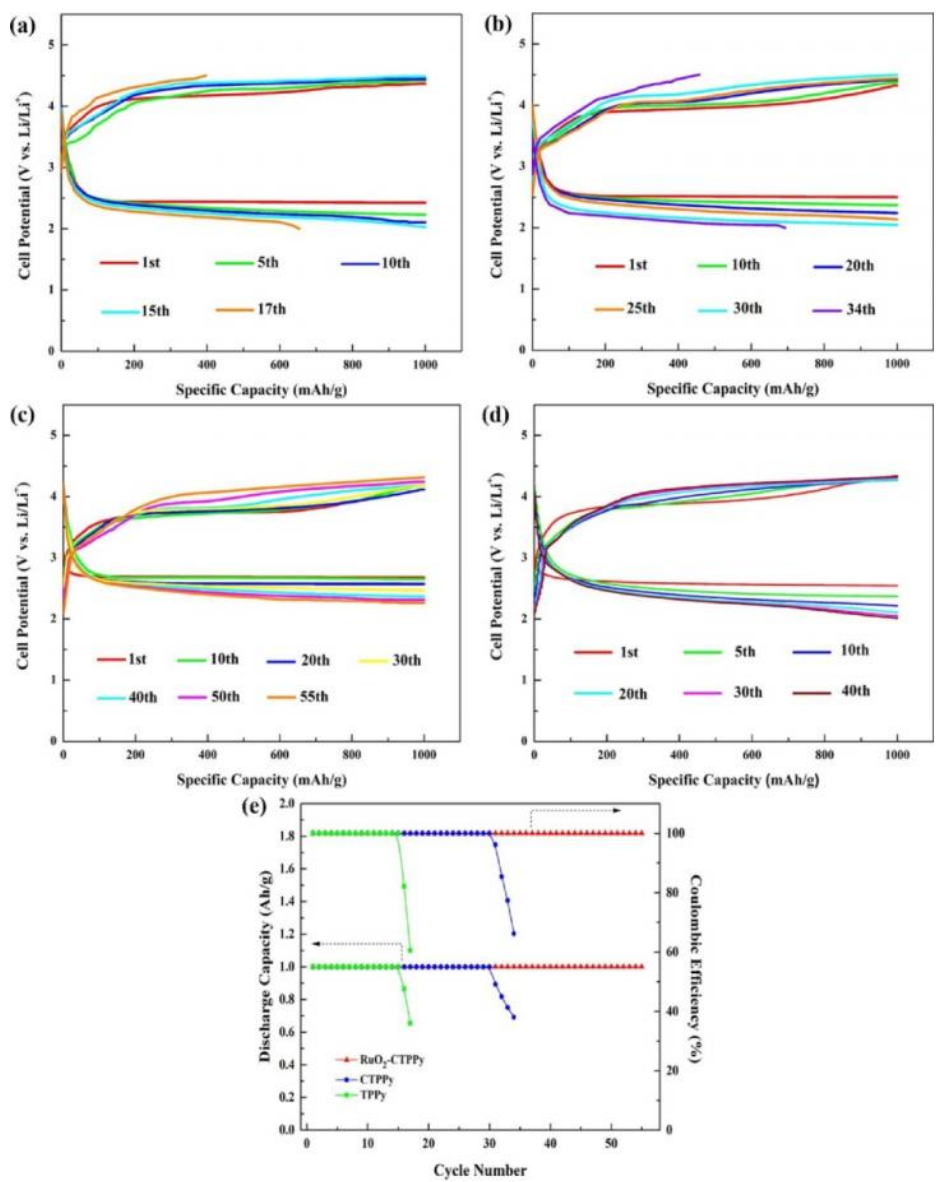

**Figure 9.** Cycling performance of a Li-$O_2$ battery at the current density of 500 mA g$^{-1}$ with a specific capacity of 1000 mAh g$^{-1}$ with the cathode made of (a) TPPy; (b) CTPPy; (c) $RuO_2$-decorated CTPPy. (d) Cycling performance of a Li-$O_2$ battery at the current density of 1000 mA g$^{-1}$ with a specific capacity of 1000 mAh g$^{-1}$ with the cathode made of $RuO_2$-decorated CTPPy. (e) The discharge capacity and coulombic efficiency of the batteries with cathode made of TPPy, CTPPy and $RuO_2$-decorated CTPPy. Reprinted from Ref. [104] with permission. Copyright (2017) Elsevier.

3.3.3. Other polymer-derived carbons

There is a number of other nanocarbons derived from different precursors that were demonstrated as electrode materials for rechargeable batteries. Like in the case of PANI and PPy derived carbons polymeric materials are largely used as sacrificial carbon sources leading to enhanced electronic transport or providing the matrix to accommodate reaction products. Here we briefly summarize some of the available reports. In Table 4 we present the reports considering Li-ion batteries, while in Table 5 the results regarding Li-S and Li-O$_2$ batteries are presented.

**Table 4.** Overview of various polymer-derived carbons used in Li-ion batteries.

| electrode material/precursor | carbonization temperature / °C | specific surface area / m$^2$ g$^{-1}$ | capacity / mA h g$^{-1}$ | reference |
|---|---|---|---|---|
| N-doped porous carbon-carbon nanotubes networks derived from melamine anchored with MoS$_2$ | | | 1218.7 @ 200 mA g$^{-1}$<br>452.2 @ 4000 mA g$^{-1}$ | [105] |
| MoO$_2$/Mo$_2$C/C microspheres polyethylene glycol | | | 665 @ 100 mA g$^{-1}$<br>588 @ 200 mA g$^{-1}$<br>(after 100 cycles) | [106] |
| electrospray method and subsequent heat treatment, using ferrous acetylacetonate, CNTs, Ketjen black, polyvinylpyrrolidone and polystyrene as raw materials | 600 | | 1317 @ 100 mA g$^{-1}$<br>after 300 cycles:<br>746 @ 1 A g$^{-1}$<br>525 @ 5 A g$^{-1}$ | [107] |
| electrospinning method in a precursor solution containing polyacrylonitrile, polystyrene, N,N-dimethylformamide and tetrahydrofuran | | | 416 | [108] |
| Waste poly(vinyl butyral) | 500 | 30.8 to 4.2 | 910 @ 0.8 A g$^{-1}$ | [109] |
| sulfonated semi-interpenetrating polystyrene Sn/C | 770 | 45.12 to 235.89 | 1210<br>331 (100 cycles) | [110] |
| Nitrogen (N) and sulfur (S) co-doped porous carbon materials polyvinyl pyrrolidone (PVP) and dodecyl sulfonate (DSO) | 800 | 1493.2 | 1175 @ 0.5 C<br>765 @ 1 C | [111] |
| Oleic acid polypropylene Fe$_3$O$_4$ organically modified montmorillonite (MMT) | 800 | 272 to 717 | 3816 @ C/5 (first)<br>410 @ C/5 (stable) | [112] |
| Si/poly(3,4-ethylenedioxythiophene):poly(styrenesulfonate) (PEDOT:PSS) | 800 | | 768 after 80 cycles | [113] |

**Table 5**. Overview of various polymer-derived carbons used in Li-S and Li-O$_2$ batteries.

**Li-S batteries**

| precursor | comment | carbonization temperature / °C | specific surface area / m$^2$ g$^{-1}$ | capacity / mA h g$^{-1}$ | reference |
|---|---|---|---|---|---|
| 2,4,6- tris(3,5-dicarboxylphenylamino)-1,3,5-triazin | N-doped carbon nanosheets pore size ~ 0.6 nm | 800 - 1000 (900 optimal) | 746.6 to 1432.6 | 1220 @ 0.1C (after 200 cycles) 727 @ 2C | [114] |
| poly(styrene-co-divinylbenzene) | | 400-1000 (900 optimal) | up to 2052 | 1066 @ 0.2 C 868 @ 0.5 C 705 @ 1C 414 @ 2C | [115] |
| (PMMA/PAN/PVP layers) formed by a co-electrospinning of three polymer solutions | KOH activation | 800 | 87.8 to 1459.2 | 1073 to 1270 @ 0.2 C | [116] |
| The cationic fluorocarbon surfactant FC4 and triblock copolymer Pluronic F127 as templates; resorcinol and formaldehyde as carbon precursors | nanospheres 80 to 400 nm mesopores 3.5 nm in diameter | 800 | 857 | 850 @ 0.09 C 400 @ 1.8 C | [117] |
| Polyacrylonitrile-S/Carbon Composite | S loading at 300 °C | 900 | 893 for carbon, reduces 10 times after loading S | 658.8 (after 50 cycles) | [118] |

**Li-O$_2$ batteries**

| precursor | comment | carbonization temperature / °C | specific surface area / m$^2$ g$^{-1}$ | capacity / mA h g$^{-1}$ | reference |
|---|---|---|---|---|---|
| carbonization using a triazine-based covalent organic polymer | up to 120 cycles with a limited capacity of 1000 mA h g$^{-1}$ | 800- 950 | 789 to 2003 | 6003 to 9960 @ 200 mA g$^{-1}$ | [119] |
| polypyrrole/cellulose composite | varying capacity by changing porosity | 1000 | 107 | 8040 @ 0.1 mA cm$^{-2}$ 480 after 3$^{rd}$ cycle | [120] |
| phenol and formaldehyde | low cycle life possible Li intercalation into cathode catalyst | 1050 | 668 to 1382 | 910 to 1852 | [121] |

## 4. Some general remarks

It is a tempting though that a deep analysis of available literature data can identify crucial properties of polymeric derived carbons that define their electrochemical behavior. However, it is seems that at this moment this is not feasible. Possible reason for such a situation might be a great variety of experimental approaches used to evaluate electrochemical performance.

Starting with electrochemical capacitors, it is interesting to observe that practically all of the presented studies evaluate capacitive performance in aqueous media. However, commercial devices use organic electrolytes and it is not unambiguously known how capacitive properties observed in aqueous media translate to the ones in organic electrolytes. While capacitive properties are always tested in concentrated electrolytes, due to acid/base character of surface functional groups capacitance will depend on the pH of the electrolyte. To make the situation even more complex, it was recently shown that capacitive response of carbon materials (tested for different graphene samples) in alkali metal chloride solutions depend on the cation present in the solution [122] and decrease from KCl to NaCl and LiCl. If one adds different electrode preparation procedures, polymer binders, addition of conductive component and so on, it becomes more understandable why it is so difficult to derive some general view on the impact of physical and chemical properties of nanocarbons on capacitive properties. However, it is clear that it is beneficial to have high concentrations of nitrogen and oxygen, while preserving high conductivity of material. Mesopores are more likely to contribute capacitance than micropores, but specific surface itself is definitely not a dominant factor in determining capacitance of polymer derived carbons. Having this in mind it is understandable why modest carbonization temperatures (600-700 °C), at least for PANI and PPy, lead to carbons with the best capacitive response. When carbonized at lower temperatures, degree of conversion to carbon is typically low and conductivity is poor. However, if carbonized at high temperatures N and O get removed from carbon surface while micropores are being developed. This results with poor hydrophylicity and low contribution of pseudocapacitance by surface functional groups and, in general, lowers capacitive response. Moreover, it is clear that it is necessary to standardize capacitive measurements to make the analysis and comparison of the results of different groups possible.

The necessity to standardize electrochemical measurements of ORR catalytic activity of carbon materials is even more obvious. While this was recognized in the case of Pt-based fuel cell catalysts [123], the order is less established in the case of carbon materials. As the apparent catalytic activity depends on the catalyst loading [50], this issue should be carefully investigated to enable proper estimate of the catalytic activity and to enable comparison of different nanocarbons. As ORR activity is measured using rotating disk electrode voltammetry, one should keep in mind that diffusion limiting current depends only on the concentration of $O_2$ and the number of electrons consumed *per* $O_2$ (for a given electrode rotation rate). As the maximum is 4 and is being reached in the case of Pt-based catalysts (in both acidic and alkaline electrolytes) it is not possible to have lower limiting current density when using Pt-based catalyst than the one measured with Pt-free nanocarbon. If this is the case, most likely Pt loading is not properly adjusted. This also leads to a problem with the comparison of catalytic activities of novel nanocarbons and benchmark Pt/C catalysts. There are additional points which refer to the measurement procedure itself. The dissolution of platinum, which is commonly used as the counted electrode, invokes necessity to abandon this practice and to use noble-metal free counter electrodes. Namely, dissolved platinum can deposit on the working electrode (noble-metal free nanocarbon) during the measurements and result with the increase of catalytic activity. Another important point is the sweep rate which is used in the measurements. In the case of Pt it is well accepted that sweep rate affects the measured catalytic activity due to the oxide place exchange process [124-126]. This leads to the decreased ORR activity when measured with lower potential sweep rates. Qualitatively the same effect was observed for PANI derived carbons [49] (Figure 10) and it was observed by the interplay of the EDL charging and charge transfer process due to the blockage of catalytically active sites. If one knows that the capacitive response, i. e. EDL charging, depends on the cation present in the solution, as mentioned, it can be anticipated that the catalytic activities of one material measured in different alkali metal hydroxides will differ. All the problems aside, the approach for making a good metal-free ORR electrocatalyst from PANI and PPy seems to be similar as in the case of fabrication of charge storage materials. Hence, good conductivity, relatively high concentrations of N and O (up to 10 at.%) and opened pore structure to allow the access of $O_2$ to the active sites are needed. There is also a prevailing opinion that pyridinic and quaternary nitrogen sites contribute ORR catalytic activity.

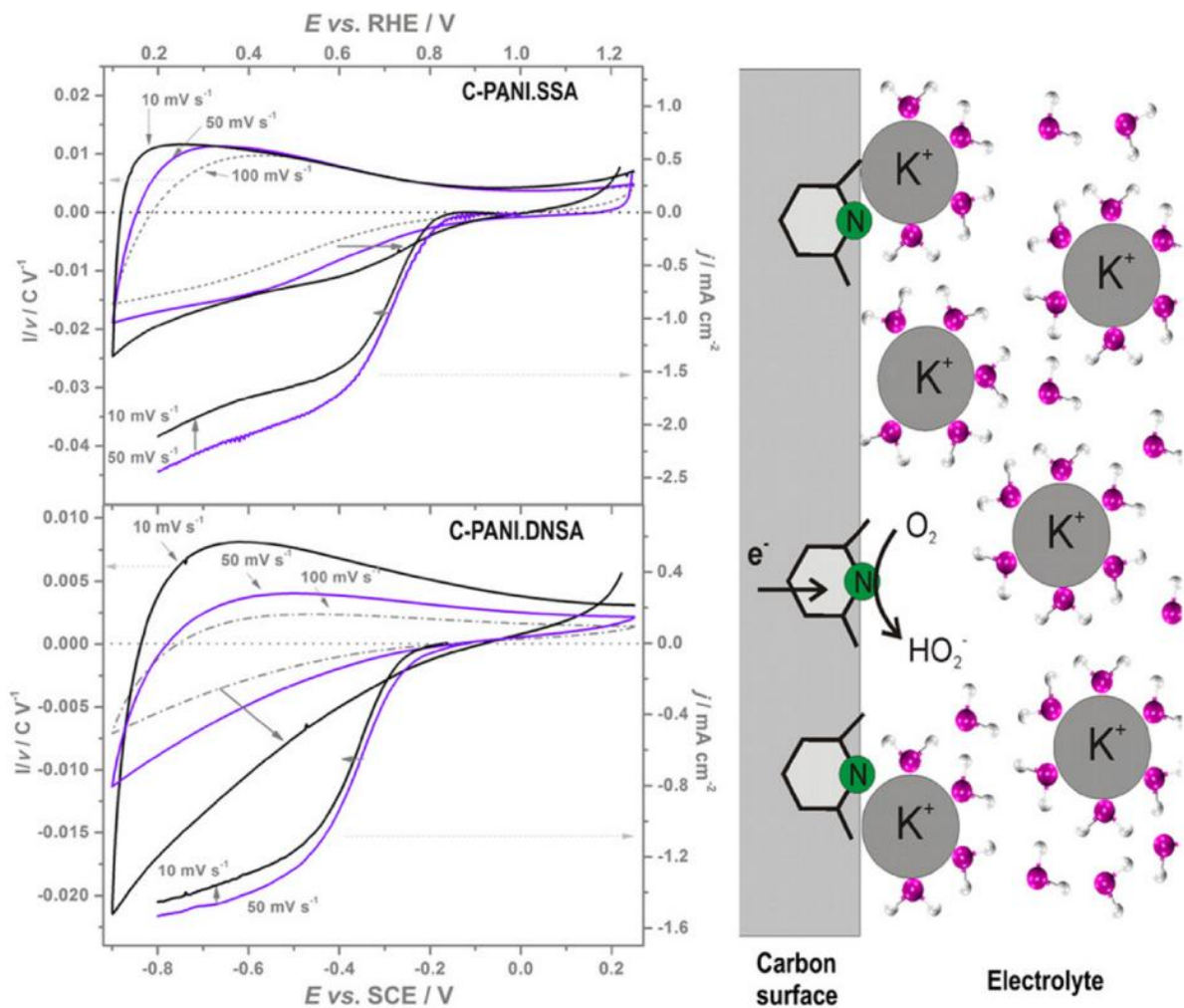

**Figure 10.** Blank cyclic voltammograms of C-PANI.SSA (top) and C-PANI.DNSA (bottom) modified GC electrode in 0.1 M KOH recorded at different sweep rates (10, 50 and 100 mV s$^{-1}$) and normalized by a matching sweep rate (I/v) and corresponding background-corrected ORR RDE currents recorded at 10 and 50 mV s$^{-1}$ (rotation rate 600 rpm, catalyst loading 250 mg cm$^{-2}$). On the right hand side a qualitative model describing blockage of ORR active sites is depicted. ORR can take place on non-blocked sites whose number is reduced if the electrode potential is swept slowly. Reprinted from Ref. [49] with permission. Copyright (2012) Elsevier.

It seems that in the case of rechargeable batteries the situation is more uniform, at least taking into consideration Li-S and Li-O$_2$ batteries where quite similar first discharge capacities are reported but materials differ cycling stability and rate capability.

In this case is it quite clear that high conductivity of produced carbon is absolutely necessary and an opened structure to accommodate products of electrode reactions. However, these are no direct correlations of battery performance with doping levels and specific surface areas of carbon materials.

Finally, we emphasize one more time that practically everything we know about carbon materials which are tested in electrochemical systems is determined *ex situ* and often under vacuum conditions, while these properties are considered to be potential independent. Very little is known about surface electrochemical process on carbon materials, at least when compared to metal electrodes, and it is mainly speculated on the basis of common knowledge of physics and chemistry.

## 5. Conclusion

There is no doubt that carbonization of polymeric precursors is an effective route for obtaining nanostructured carbons with desired morphology and proper chemical and physical properties needed for the application in electrochemical power sources. This approach not only allows for production of high performance carbons, which can be used in supercapacitors, as electrocatalysts or battery materials, but also provides versatile route for production of different composite materials for energy conversion and storage applications. Nanocarbons derived from PANI and PPy were in the focus of this work and their use allows reaching superior electrochemical performance of different electrochemical power sources. However, it is clear that more systematic analysis of properties-performance links is needed with the introduction of novel approaches to investigate carbons directly under operating conditions.

**Acknowledgement**

The financial support by Ministry of Education, Science and Technological Development of the Republic of Serbia (projects III45014 and OI172043) is highly appreciated.

**Acknowledgement**

The financial support by Ministry of Education, Science and Technological Development of the Republic of Serbia (projects III45014 and OI172043) is highly appreciated.